\documentclass[aps,prl,preprint,showpacs]{revtex4}
\usepackage{bm}
\begin{document}

\title{Superstatistics as the statistics of quasi-equilibrium states: Application to fully developed turbulence}

\author{Mohammad Mehrafarin}
\email{mehrafar@aut.ac.ir} 
\affiliation{Physics Department, Amirkabir University of Technology, Tehran
15914, Iran}

\begin{abstract}
In non-equilibrium states, currents are produced by irreversible processes that take a system toward the equilibrium state, where the current vanishes. We demonstrate, in a general setting, that a superstatistics arises when the system relaxes to a (stationary) quasi-equilibrium state instead, where only the \textit{mean} current vanishes because of fluctuations. In particular, we show that a current with Gaussian white noise takes the system to a unique class of quasi-equilibrium states, where the superstatistics coincides with Tsallis escort $q$-distributions. Considering the fully developed turbulence as an example of such quasi-equilibrium states, we analytically deduce the power-law spectrum of the velocity structure functions, yielding a correction to the log-normal model which removes its shortcomings with regard to the decreasing higher order moments and the Novikov inequality, and obtain exponents that agree well with the experimental data.
\end{abstract}


\maketitle

\section{Introduction}
Recent experimental and numerical studies have shown that many complex systems, and of particular relevance to this work, the fully developed turbulent system \cite{Beck,Arimitsu,Beck1,Beck2,Beck3,Beck4,Gotoh},  exhibit non-Gibbsian distributions that are well fitted by Tsallis $q$-distributions \cite{Tsallis}. When these distributions are involved, physical mean values correspond to the $q$-average values, which are averages over the postulated escort $q$-distributions.  The physical interpretation of the escort distributions is, however, not clear. Addressing from a physical foundation the question of why such generalized distributions arise is crucial to understanding the behavior of these complex systems. Despite several attempts \cite{Beck2,Wilk,Beck5,Chavanis}, a comprehensive answer to this question is still lacking. In this report, we approach the problem from a general perspective that provides a new insight and consider, as an application, the example of fully developed turbulence.

We consider a macroscopic system with a relevant extensive variable $X$ and a conjugated (entropic) intensive variable $P=\partial_X S$, where $S(X,\ldots)$ is the system's entropy. Let $X_0$, $P_0$, $S_0$ represent the same quantities for the system's environment, which acts as a reservoir. The closed composite system has entropy $S_{\text{tot}}=S+S_0$, and the closure condition requires that $X+X_0=X_{\text{tot}}\ (\text{const.})$. If $X$ and $X_0$ are otherwise unconstrained, their equilibrium values $\tilde{X}$ and $X_{\text{tot}}-\tilde{X}$ are obtained by the vanishing of the affinity \cite{Callen} (the thermodynamic force) $F=\partial_X S_{\text{tot}}=P-P_0$. Thus, if $F$ is zero the system is in equilibrium, but if $F$ is nonzero an irreversible process occurs, producing a current $\dot{X}$ that takes the system toward the equilibrium state (dot denotes time derivative). Since the current vanishes if the affinity vanishes, we can expand the current in terms of the affinity with no constant term. Assuming sufficiently small affinities throughout, we have $\dot{X}=\ell F$, where $\ell>0$ is called the Onsager kinetic coefficient. This relationship between the current and the affinity characterizes the relaxation flow of the system. Thus, since
\begin{equation}
\dot{F}=(\partial_X P)\; \ell F\approx \partial_X P|_{\tilde{X}}\; \ell F \label{relxn}
\end{equation}
the equilibrium state, $F=0$, is an attractor provided $\partial_X P|_{\tilde{X}}=\partial_X^2 S|_{\tilde{X}}<0$. This is, of course, the stability criterion which implies that the response function (like the heat capacity, the compressibility, and so on) is positive. 

In the equilibrium state, the probability distribution function (p.d.f.) of the intensive variable, therefore, is $\rho_0(P)=\delta(P-P_0)$. The extensive variable is, however, distributed about the mean equilibrium value $\tilde{X}$ with a p.d.f. $p_0(X)\propto w(X) e^{-P_0X}$ ($k_B=1$), where $w(X)$ is the `density of states'. This standard result, which forms the basis of of equilibrium statistical mechanics, is obtained by applying the principle of equal probability of microstates to the closed composite system and imposing the equilibrium condition $P=P_0$. The latter condition marginalizes the $P$-values, implying that the equilibrium distribution, $p_0(X)$, is in fact a marginal p.d.f. If, instead of the equilibrium state, the current takes the system to a different attractor where the intensive variable is distributed about $P_0$, we will have a stationary (a quasi- or meta-equilibrium) state whose $P$-p.d.f., $\rho(P)$, must be used to marginalize the $P$-values. In such quasi-equilibrium states, the intensive variable fluctuates about the mean equilibrium value so that in these states only the \textit{mean} current is zero. This suggests that a fluctuating current (due to the underlying noise generating mechanisms) drives the system to such quasi-equilibrium states, where it attains zero mean value. For simplicity we shall consider systems with a constant density of states so that $p_0(X)=P_0 e^{-P_0X}$ and $\langle X\rangle_0=\tilde{X}=1/P_0$. (The formulation can be easily extended to the more general case of $w(X)\propto X^{\alpha-1}$ ($\alpha>0$), where our central result in the last section, equation (\ref{stat}), is similarly shown to follow just the same). The (marginal) quasi-equilibrium $X$-p.d.f. would then be given by the Bayes rule according to
\begin{equation}
p(X)= \int_0^\infty Pe^{-PX} \rho(P) dP=\int_0^\infty e^{-PX} \rho(\ln \frac{P}{P_0}) dP \label{super}
\end{equation}
which reduces to the equilibrium $X$-p.d.f. $p_0(X)$ if $\rho(P)=\rho_0(P)=\delta(P-P_0)$, as it should.

Generalized distributions of the form (\ref{super}), which have been called superstatistics \cite{Beck5,Cohen}, have found many applications in diverse physical problems \cite{Beck4,Chavanis,Sattin,Ausloos,Beck6}. From our perspective, therefore, a superstatistics is the most natural extension of the equilibrium Gibbs-Boltzmann statistics to (stationary) quasi-equilibrium sates, where only the mean current vanishes because of fluctuations. In complex systems, we can assume that these fluctuations arise from the interaction of a very large number of independent microscopic degrees of freedom with identical stochastic properties. The central limit theorem then implies that the fluctuations are Gaussian distributed. Naturally assuming that they are temporally uncorrelated too, we have a Gaussian white noise that contains all frequencies.

In the sequel we show that adding a Gaussian white noise to the current changes the attractor and yields a new attractor state with $\rho(P)$ as the $\Gamma$-distribution, resulting in the Tsallis escort q-distributions as the superstatistics. In these states, the affinity and the current are Gaussian distributed with zero mean. In other words, we show that a current with Gaussian white noise fluctuations takes the system to a unique class of quasi-equilibrium states, where it attains the ubiquitous Gaussian distribution with zero mean and the superstatistics coincides with the Tsallis escort $q$-distributions. This explains the ubiquity and the physical significance of escort distributions and, hence, of the $q$-averages as quasi-equilibrium mean values. 

The noise maybe generated, in externally driven complex systems for example, by the variation of a control parameter. The resulting change in the attractor and the appearance of a new quasi-equilibrium behavior associated with the superstatistics is then an instance of bifurcation. Regarding turbulent flow as such, the control parameter is the Reynolds number and/or the boundary roughness \cite{Goldenfeld,Mehrafarin}, the relevant extensive variable is the kinetic energy (per unit mass) associated with a given spatial scale, and the current is its rate of change per unit time due to dissipation. The current, being scale-dependent, fluctuates in a scale-dependent manner (intermittency) as the value of the control parameter is increased, driving the system to a quasi-equilibrium state where the mean dissipation vanishes at intermediate scales. The new behavior associated with this quasi-equilibrium state, which is identified with the fully developed turbulence state, is the power-law behavior of the velocity structure functions. We shall analytically deduce the power-law spectrum, yielding a correction to the log-normal model which removes its shortcomings with regard to the decreasing higher order moments and the Novikov inequality \cite{Novikov}, and obtain exponents that agree well with the available data.

\section{Fluctuating current: relaxation to quasi-equilibrium state}
Introducing the response function $C_0=-\partial_P X|_{P_0}>0$, we write equation (\ref{relxn}) as
\begin{equation}
\dot{F}=-\frac{1}{\tau} F \label{F}
\end{equation}
where $\tau=C_0/\ell$ is the relaxation time of the system. The unique stationary solution of this equation is $F=0$, which corresponds to the equilibrium state at $t\gg \tau$. Equation (\ref{F}) can be cast as an equation for $P$ instead. Writing $\frac{P}{P_0}=1+\frac{F}{P_0}$, we have
\begin{equation}
\frac{d}{dt}\ln \frac{P}{P_0}=-\frac{1}{\tau} \frac{F}{P_0}\left[1-\frac{F}{P_0}+(\frac{F}{P_0})^2-\cdots\right]=-\frac{1}{\tau}(\frac{P}{P_0}-1) \label{P}
\end{equation}
to first order in the (small) affinity. Adding the Gaussian white noise, $\xi(t)$, with the properties
\begin{equation}
\langle \xi(t)\rangle=0, \ \ \langle \xi(t)\: \xi(t^\prime)\rangle=2D\delta(t-t^\prime) \label{noise}
\end{equation}
($D>0$) to the current $\dot{X}=\ell F$ is equivalent to adding it to $F$ (with $D$ in (\ref{noise}) redefined), which, in turn, is equivalent to adding it to the right hand side of equation (\ref{P}) (with $D$ redefined again). We, thus, arrive at the following stochastic differential equation for $P$:  
\begin{equation}
\frac{d}{dt}\ln \frac{P}{P_0}=-\frac{1}{\tau}(\frac{P}{P_0}-1)+\xi(t). \label{P2}
\end{equation}
This equation yield a new attractor (stationary) state, as we now show. To this end, it is more convenient to cast (\ref{P2}) in the form of a standard Langevin equation by introducing $v=\ln \frac{P}{P_0}$. We have
$\dot{v}=f(v)+\xi(t)$ where $f(v)=-\frac{1}{\tau}(e^v-1)$. The corresponding Fokker-Plank equation is \cite{Kampen} $\partial_t \rho(v,t)+\partial_v j(v,t)=0$ where $j(v,t)=f(v)\rho(v,t)-D\partial_v \rho(v,t)$.
The stationary solution, $\rho(v)$, at $t\gg \tau$ satisfies $j=0$, yielding
\begin{equation}
\rho(\ln \frac{P}{P_0})= [\Gamma(\frac{1}{q-1})]^{-1}\:(\frac{1}{q-1}\frac{P}{P_0})^{\frac{1}{q-1}}\: \exp (-\frac{1}{q-1}\frac{P}{P_0}) \label{gamma}
\end{equation}
where, $q-1=D\tau>0$. The P-p.d.f. $\rho(P)=\frac{1}{P}\rho(\ln \frac{P}{P_0})$ is, therefore, the well known $\Gamma$-distribution. Its mean is $P_0$ and its variance, $(q-1)P_0^2$, vanishes as $D\rightarrow 0$ ($q\rightarrow 1$), i.e. as the fluctuations disappear, recovering the p.d.f. $\delta(P-P_0)$ of the equilibrium state. This identifies the noise as the sole agent of driving the system toward the quasi-equilibrium state. Note also, by similarly writing the standard Langevin equation (\ref{F}) (with the noise added, of course) as a Fokker-Plank equation, that the quasi-equilibrium affinity, and hence the current, are Gaussian distributed about zero, approaching the delta function as the fluctuations disappear. Substituting (\ref{gamma}) in (\ref{super}), after a simple integration we obtain the quasi-equilibrium superstatistics as $p(X)=P_0\:\left[1+(q-1)P_0X\right]^{-\frac{q}{q-1}}$, which coincides with the Tsallis escort $q$-distribution, $p^{(q)}(X)$. Of course, $q=1+D\tau\geq 1$, where for $q=1$ the distribution reduces to the equilibrium exponential distribution. Note that the escort distribution itself has emerged as the physically relevant p.d.f. and is not postulated via the direct Tsallis q-distribution. Consequently, the quasi-equilibrium mean value $\langle X\rangle=\int_0^\infty Xp(X) dX=\frac{1}{(2-q)P_0}$, which exists provided $q< 2$ (thus, $1\leq q<2$), is inherently a $q$-average.  Finally, introducing the normalized variable $x=X/\langle X\rangle$, we write
\begin{equation}
p(x)=\frac{1}{2-q}\left[1+\left(\frac{q-1}{2-q}\right)x\right]^{-\frac{q}{q-1}} \label{escort}
\end{equation}
and note that $X$ may be taken to denote an extensive quantity per unit mass, when appropriate. (We may note, in passing, that the physical significance of the extensivity parameter, $q$, coincides with interpretation given in reference \cite{Wilk}. There, the authors obtain the direct q-distributions by considering fluctuations of the parameters of the usual exponential distribution. Their postulated Langevin equation (11), which yields the $\Gamma$-distribution, coincides with our equation (\ref{P2}), provided we make the identification $\chi=1/P$ and $\phi=1/\tau P_0$. However, the integrand of their integral expression (4) slightly differs from ours in equation (\ref{super}).)

\section {Fully developed turbulence as a quasi-equilibrium state}
In a turbulent flow, the current responsible for relaxation is the rate of change of the kinetic energy per unit mass associated with a given spatial scale. This is because of dissipation, which causes a difference between the in-flux from the larger scales and the out-flux to the smaller ones. For sufficiently high values of the control parameter (Reynolds number and/or boundary roughness), the system relaxes to a stationary state where the mean in- and out-fluxes are equal at finite intermediate scales and dissipation occurs at a mean rate $\varepsilon$ only at the smallest (Kolmogorov) scale. This state, which is the state of vanishing mean current, corresponds to the quasi-equilibrium state of fully developed turbulence.

Let $u(r)$ denote the longitudinal velocity difference in a turbulent flow as measured on an intermediate length scale $r$. In the simplest model, we can take $\frac{1}{2}u^2$ as a measure of the kinetic energy (per unit mass) of the flow associated with scale $r$. We, therefore, take $X(r)=\frac{1}{2}u^2(r)$ as the relevant (scale-dependent) extensive quantity (per unit mass) with its rate of change, $\dot{X}(r)$, as the (scale-dependent) current. Were intermittent fluctuations not present in the current, the turbulent system would relax to the equilibrium state with mean energy $\tilde{X}(r)=\frac{1}{2}\langle u^2(r)\rangle_0$, and the fully developed turbulent state would be an equilibrium state. The normalized variable is, thus, $x(r)=u^2/\langle u^2\rangle_0$ and the equilibrium p.d.f. $p_0(x)=e^{-x}$ determines the distribution of $|u|$ ($-\infty<u<\infty$) according to $p_0(|u|)=\frac{2|u|}{\langle u^2\rangle_0}\: e^{-|u|^2/\langle u^2\rangle_0}$.
This is the p.d.f for u irrespective of its sign, i.e. $p_0(|u|)=p_0(u)+p_0(-u)$, and so the even part of the $u$-p.d.f., $p_0(u)$, is only obtained; the odd part (which must satisfy the exact constraint imposed by the Kolmogorov's four-fifths law \cite{Kolmogorov} for the third moment) remains undetermined. This implies that we can obtain only the even moments of the velocity differences. However, the moments are sometimes defined in terms of $|u|$ instead of $u$, in which case the even moments coincide. Although there has been no theoretical justification for the spectrum of the odd moments thus obtained, experimental and numerical evidence suggest that the spectrum does not change significantly \cite{Benzi,Stolovitzky,Benzi2,Arneodo}. In particular, the third moment of $|u|$ has a scaling exponent that is close to unity, which is curious since the four-fifths law pertains to $u$ and not $|u|$. This has motivated the very useful notion of extended self-similarity (ESS) \cite{Benzi,Benzi2}, according to which moments are conveniently plotted against the third moment as a log-log plot.  As shall see, using the absolute velocity differences, our theory drives an exponent very close to unity for the third moment, thus providing an explanation for the ESS. Hence ($m=1,2,\ldots$)
\begin{equation}
\langle |u|^m\rangle_0=\int_{0} ^\infty |u|^m p_0(|u|)d|u|=\Gamma(\frac{m}{2}+1)\:\langle u^2\rangle_0^{m/2}. \label{kol}
\end{equation}
Since current fluctuations are ignored, the free parameter of the p.d.f. can be determined $\grave{\text{a}}$ la Kolmogorov \cite{Kolmogorov2} as $\langle u^2\rangle_0 \sim(\varepsilon r)^{2/3}$, to leading order. Then $\langle |u|^m\rangle_0\sim (\varepsilon r)^{m/3}$, which is the expected Kolmogorov power-law spectrum. Therefore, the Kolmogorov spectrum would apply if the fully developed turbulent flow were an equilibrium phenomenon.

Let us now consider adding a Gaussian white noise, which is anticipated at sufficiently high values of the control parameter, to the current $\dot{X}(r)$. Since the current is scale dependent, we consider scale-dependent fluctuations by taking $D=D(r)$ so that $q(r)=1+D(r)\tau$ now depends on the scale. In the presence of such intermittent fluctuations, the system relaxes in the quasi-equilibrium state of fully developed turbulence where the mean dissipation vanishes (at intermediate scales) yielding zero mean current. The quasi-equilibrium p.d.f. is the (scale-dependent) escort distribution (\ref{escort}), which similarly implies  
\[
p(|u|)=\frac{2|u|}{(2-q)\langle u^2\rangle}\left[1+\left(\frac{q-1}{2-q}\right)\frac{u^2}{\langle u^2\rangle}\right]^{-\frac{q}{q-1}}. 
\]
It has two free parameters, namely, $q$ and $\langle u^2\rangle$. The moments are, hence, given by
\[
\langle |u|^m\rangle=\int_0 ^\infty |u|^m p(|u|)d|u|=\frac{(2-q)^{m/2}\langle u^2\rangle^{m/2}}{(q-1)^{\frac{m}{2}+1}}\int_0^1 dt\: t^{\frac{1}{q-1}-\frac{m}{2}-1} (1-t)^{m/2}.
\]
The integral exists provided $m<2(q-1)^{-1}$, which includes sufficiently large $m$'s when $q$ is close to unity. We assume this is the case (see shortly). Then 
\begin{equation}
\frac{1}{\Gamma(\frac{m}{2}+1)}\frac{\langle |u|^m\rangle}{\langle u^2\rangle^{m/2}}= \frac{\Gamma(\frac{1}{q-1}-\frac{m}{2})}{\Gamma(\frac{1}{q-1})}\left(\frac{2-q}{q-1}\right)^{m/2}\label{stat} 
\end{equation}
which reduces to (\ref{kol}) for $q=1$, of course. This will give the velocity statistics in the fully developed turbulence state. Our aim is to drive the structure functions in the form
\begin{equation}
\langle |u|^m\rangle \sim (\varepsilon r)^{m/3} (\frac{r}{L})^{\eta_m} \label{interm}
\end{equation}
to leading order, where $L(\gg r)$ is the integral length scale and the proportionality constants are non-universal, depending on the detailed geometry of the turbulent production. 

The free parameter, $\langle u^2\rangle$, is to be determined by (\ref{interm}) with $m=2$, so that
\begin{equation}
\langle u^2\rangle \sim (\varepsilon r)^{2/3} (\frac{r}{L})^{\eta_2}. \label{u2}
\end{equation}
From (\ref{stat}) we have
\[
\frac{\langle |u|^m\rangle}{\langle |u|^{m^\prime}\rangle}=\frac{\Gamma(\frac{m}{2}+1)}{\Gamma(\frac{m^\prime}{2}+1)} \langle u^2\rangle^{\frac{m-m^\prime}{2}}\left(\frac{2-q}{q-1}\right)^{\frac{m-m^\prime}{2}} \frac{\Gamma(\frac{1}{q-1}-\frac{m}{2})}{\Gamma(\frac{1}{q-1}-\frac{m^\prime}{2})}
\]
which, for $m> m^\prime$ both odd/even, becomes
\begin{equation}
\frac{\langle |u|^m\rangle}{\langle |u|^{m^\prime}\rangle}= \frac{\Gamma(\frac{m}{2}+1)}{\Gamma(\frac{m^\prime}{2}+1)} \langle u^2\rangle^{\frac{m-m^\prime}{2}}(2-q)^{\frac{m-m^\prime}{2}}\prod_{l=1}^{\frac{m-m^\prime}{2}} [1-(l+\frac{m^\prime}{2})(q-1)]^{-1}. \label{a1}
\end{equation}
Since $(2-q)^{-1}=q+O((q-1)^2)=q$ to leading order, and the product is
\begin{eqnarray}
\exp \left(-\sum_{l=1}^{\frac{m-m^\prime}{2}} \ln [1-(l+\frac{m^\prime}{2})(q-1)]\right)=\nonumber \ \ \ \ \ \ \ \ \ \ \ \ \ \ \ \ \ \ \ \ \ \ \ \ \\
\exp \left([(\frac{m^2}{2}+m)-(\frac{m^{\prime 2}}{2}+m^\prime)]\frac{q-1}{4}+O((q-1)^2)\right)  
=q^{[(\frac{m^2}{2}+m)-(\frac{m^{\prime 2}}{2}+m^\prime)]/4} \nonumber
\end{eqnarray} 
having used $\ln q=(q-1)+O((q-1)^2)$, (\ref{a1}) yields 
\begin{equation}
\langle |u|^m\rangle=\Gamma(\frac{m}{2}+1)\langle u^2\rangle^{\frac{m}{2}} q^{m(m-2)/8} \label{er}
\end{equation}
which checks for $m=2$ and reduces to (\ref{kol}) for $q=1$, as it should. For $m=4$, we have $\langle u^4\rangle/\langle u^{2}\rangle^2= 2q$, so that the (scale dependent) value of the kurtosis (flatness) determines the other free parameter, $q(r)$. Using (\ref{interm}), we find 
\begin{equation}
q(r) \sim (\frac{r}{L})^{-(2\eta_2-\eta_4)}. \label{q}
\end{equation}
Since $2\eta_2-\eta_4 \ll 1$, $q(r)$ is a slowly varying function of $r$ in the intermediate scales and, hence, so is the upper bound $m_{\text{max}}(r)=2(q-1)^{-1}$ on $m$. Although the reliability of experimental data rapidly decreases for moments of order higher than $m=10$, a scale-dependent upper bound on the order is at variance with the general supposition that moments of all orders exist. However, as we shall see below, it will serve to remove the shortcomings of the log-normal model with regard to its decreasing higher order moments and the Novikov inequality \cite{Novikov}. Substituting in (\ref{er}) our expressions for $\langle u^2\rangle$ and $q$ from (\ref{u2}) and (\ref{q}), we finally obtain the spectrum (\ref{interm}) with the intermittency exponents  
\begin{equation}
\eta_m=-\frac{1}{2}m(m-3)\eta_2+\frac{1}{8}m(m-2)(2\eta_2+\eta_4).\label{exp}
\end{equation} 

The first term coincides with the prediction of the log-normal (LN) model \cite{Obukhov} (with $\eta_2\equiv \mu/9$ as the free parameter), which is known to yield good agreement with experimental data for moments up to order $m=10$. Indeed, if we substitute the LN value $\eta_4=-2\eta_2$ in (\ref{exp}), then the second term vanishes. Thus, the second term represents deviation from log-normality and yields corrections to the LN predictions. Denoting the relative deviation of $\eta_4$ from its LN value by $\delta$ ($|\delta|< 1$), (\ref{exp}) can be written as 
\begin{equation}
\eta_m=\eta_m^{\text{LN}}-\frac{1}{4}m(m-2)\eta_2 \delta. \label{eta1}
\end{equation}
Writing (\ref{interm}) as $\langle |u|^m\rangle \sim r^{\zeta_m}$, the scaling exponents $\zeta_m=m/3+\eta_m$ of our `corrected' LN model are, therefore,
\begin{equation}
\zeta_m=\zeta_m^{\text{LN}}-\frac{1}{4}m(m-2)\eta_2 \delta. \label{zeta} 
\end{equation}
Since the LN exponents for higher moments generally fall below experimentally measured values, we anticipate a negative $\delta$ (for example, $\delta \approx -0.08$, as given by the She-Leveque model \cite{She}). For $m>m_0^{\text{LN}}=\frac{3}{2}+\frac{1}{3\eta_2}(\approx 16.5)$ (for numerical estimates we use the LN value, $\eta_2\approx 0.2/9$), the scaling exponents $\zeta_m^{\text{LN}}$ monotonically decrease with $m$, implying the breakdown of the incompressibility approximation in the fully developed turbulent state \cite{Frisch}. From (\ref{zeta}), $\zeta_m$ is a monotonically increasing function of $m$ provided $m<m_0=\frac{2m_0^{\text{LN}}+\delta}{2+\delta}$, and for $\delta<0$ we have $m_0>m_0^{\text{LN}}$. Therefore, the exponents $\zeta_m$ increase with $m$ if the slowly varying upper bound $m_{\text{max}}(r)$ remains sufficiently close to $m_0^{\text{LN}}$. That is, if $q-1$ remains close to $2/m_0^{\text{LN}}(\approx 0.12)$, which is consistent with our assumption that $q-1\ll 1$. In other words, for $|q(r)-1-\frac{2}{m_0^{\text{LN}}}|\ll 1$, the corrected model (with $\delta<0$) rids the LN model of its decreasing higher order moments, yielding structure functions only up to order $[m_0^{\text{LN}}]$ ($[\:]$ stands for the integer part). Then, it will also satisfy the Novikov inequality $\eta_m+m > 0$, which is violated by the LN model. This can be readily checked using (\ref{eta1}) and noting that $\frac{6-\delta}{10-\delta}<\frac{m_0^{\text{LN}}}{2}<\frac{6 m_0^{\text{LN}}+\delta-6}{2+\delta}$.

Using the tentative values $\eta_2= 0.2/9$ and $\delta=-0.08$, Table \ref{tab1} compares $\zeta_m$ up to order $m=10$, as given by our theory (equation (\ref{zeta})), the LN model and the references \cite{Benzi2,Anselmet}. The agreement is quite good considering the uncertainties involved. In particular, we notice the value of $\zeta_3$, which is only slightly larger than the exact value 1 (implied by the four-fifths law for $\langle u^3 \rangle$) in conformance with $\langle |u|^3 \rangle>|\langle u^3 \rangle|$ and the data of \cite{Benzi2}, theoretically justifying ESS. 
\begin{table}
\caption{Comparison of $\zeta_m$ as given by our theory, the LN model, and experiment.}
\begin{ruledtabular}
\begin{tabular}{ccccc}
$m$ & Theory & LN Model& Ref. \cite{Benzi2}& Ref.\footnote{Moments pertain to $u$ and so $\zeta_3=1$ by assumption.}\cite{Anselmet}\\
\hline
1&0.36&0.36&0.37&-\\
2&0.69&0.69&0.70&0.71\\
3&1.00&1&1.00&1\\
4&1.29&1.29&1.28&1.33\\
5&1.56&1.56&1.54&1.54\\
6&1.81&1.8&1.78&1.8\\
7&2.04&2.02&2.00&2.06\\
8&2.24&2.22&2.23&2.28\\
9&2.43&2.4&-&2.41\\
10&2.59&2.56&-&2.60\\
\end{tabular}
\end{ruledtabular}	
\label{tab1}
\end{table}

\end{document}